# The influence of the in-plane lattice constant on the superconducting transition temperature of $FeSe_{0.7}Te_{0.3}$ thin films

Feifei Yuan[1,2], Kazumasa Iida[3, a)], Vadim Grinenko[2], Paul Chekhonin[2,4], Aurimas Pukenas[4], Werner Skrotzki[4], Masahito Sakoda[5], Michio Naito[5], Alberto Sala[6], Marina Putti[6], Aichi Yamashita[7], Yoshihiko Takano[7], Zhixiang Shi[1, a)], Kornelius Nielsch[2] and Ruben Hühne[2, a)]

[1]*Department of Physics and Key Laboratory of MEMS of the Ministry of Education, Southeast University, Nanjing 211189, People's Republic of China*

[2]*Institute for Metallic Materials, IFW Dresden, D-01069 Dresden, Germany*

[3]*Department of Materials Physics, Graduate School of Engineering, Nagoya University, Nagoya 464-8603, Japan*

[4]*Institute of Structural Physics, Dresden University of Technology, 01062 Dresden, Germany.*

[5]*Department of Applied Physics, Tokyo University of Agriculture and Technology, Koganei, Tokyo 184-8588, Japan*

[6]*Dipartimento di Fisica, Università di Genova and CNR-SPIN, Via Dodecaneso 33, I-16146 Genova, Italy*

[7]*National Institute for Materials Science (NIMS), Tsukuba, Ibaraki, 305-0047, Japan*

Epitaxial Fe(Se,Te) thin films were prepared by pulsed laser deposition on $(La_{0.18}Sr_{0.82})(Al_{0.59}Ta_{0.41})O_3$ (LSAT), $CaF_2$-buffered LSAT and bare $CaF_2$ substrates, which exhibit an almost identical in-plane lattice parameter. The composition of all Fe(Se,Te) films were determined to be $FeSe_{0.7}Te_{0.3}$ by energy dispersive X-ray spectroscopy, irrespective of the substrate. Albeit the lattice parameters of all templates have comparable values, the in-plane lattice parameter of the $FeSe_{0.7}Te_{0.3}$ films varies significantly. We found that the superconducting transition temperature ($T_c$) of $FeSe_{0.7}Te_{0.3}$ thin films is strongly correlated with their $a$-axis lattice parameter. The highest $T_c$ of over 19 K was observed for the film on bare $CaF_2$ substrate, which is related to unexpectedly large in-plane compressive strain originating mostly from the thermal expansion mismatch between the $FeSe_{0.7}Te_{0.3}$ film and the substrate.

Iron chalcogenides are attractive materials due to their very high superconducting transition temperature ($T_c$) above 100 K in the form of monolayer[1, 2]. Additionally, iron chalcogenides have the simplest crystal structure among the Fe-based superconductors (FBS) and the less toxic nature, which is favorable for fundamental studies as well as applications[3-6]. In the case of $FeSe_{1-x}Te_x$ thin films, $T_c$ varies considerably depending on growth condition and substrate material[7-15]. However, the reason for such a large $T_c$ variation in thin films is still under debate. Bellingeri *et al.* suggested that the lattice parameters and the superconducting properties of Fe(Se,Te) thin films show a non-trivial dependence on the in-plane lattice constant of the substrates on which they are deposited[7]. Fe(Se,Te) thin films with high crystalline quality as well as excellent superconducting properties have been prepared on $CaF_2(001)$[8, 9], $CeO_2$-buffered single crystals (Y-stabilized $ZrO_2$ and $SrTiO_3$)[10, 11], and

---

[a)] Authors to whom correspondence should be addressed. Electronic addresses: zxshi@seu.edu.cn , iida@mp.pse.nagoya-u.ac.jp and R.Huehne@ifw-dresden.de.

technical substrates due to the good matching of the *a*-axis length between the various templates and Fe(Se,Te) film ($a$ = 3.801 Å)[16], where the respective lattice parameters of $CaF_2$ and $CeO_2$ are $a/\sqrt{2}$ = 3.862 Å and $a/\sqrt{2}$ = 3.82 Å[8,10]. However, it was shown that the lattice parameter *a* of Fe(Se,Te) thin films on $CaF_2$ is shorter than that of bulk materials[7-9, 12-15], whereas a larger value would be expected for a coherent growth on this substrate. On the other hand, $(La_{0.18}Sr_{0.82})(Al_{0.59}Ta_{0.41})O_3$ (LSAT) seems to be a good template, as the lattice parameter *a* of LSAT is 3.868 Å, which is almost the same value as that of the $CaF_2$ substrate, resulting in a lattice mismatch of -1.76% for $FeSe_{0.5}Te_{0.5}$ films[16]. However, Imai *et al.* did not observe epitaxial growth of Fe(Se,Te) on LSAT(001) and additionally no superconductivity[17-19]. Hence, the question arises, why Fe(Se,Te) films on $CaF_2$ and LSAT substrates show contrasting crystalline and superconducting properties, even though the in-plane lattice parameter of $CaF_2$ is almost identical to that of LSAT. In this letter, $FeSe_{0.7}Te_{0.3}$ thin films were deposited on three different templates with similar in-plane lattice constant, i.e. bare LSAT, $CaF_2$-buffered LSAT and bare $CaF_2$ single crystal substrates in order to investigate the key factor for achieving high $T_c$. We show that the in-plane lattice parameter for $FeSe_{0.7}Te_{0.3}$ strongly depends on substrate and correlates with the superconducting transition temperature. The origin of this behavior is discussed in terms of interface properties and thermal expansion mismatch.

$CaF_2$ buffer layers with different thickness between 20 nm and 80 nm were deposited on LSAT(001) substrates in a customer-designed molecular beam epitaxy (MBE) chamber at 400℃. Afterwards, Fe(Se,Te) films were deposited on the aforementioned templates by pulsed laser deposition (PLD) with a KrF excimer laser (wavelength: 248 nm, repetition rate: 7 Hz) under UHV conditions with a background pressure of $10^{-9}$ mbar as described in our previous report[14]. The nominal $FeSe_{0.5}Te_{0.5}$ target prepared by a melting process was used for the deposition process. The detailed target preparation was found in Ref. 20. The substrate temperature was fixed at 360℃ during deposition.

The structural properties of the films were investigated by X-ray diffraction (XRD) in $\theta$-$2\theta$ geometry at a Bruker D8 Advance with Co-$K_\alpha$ radiation (wavelength: 0.178896 nm) in order to reduce the fluorescence and at a texture goniometer Philips X'Pert with Cu-$K_\alpha$ radiation (wavelength: 0.15418 nm). The *c*-axis lattice parameters of Fe(Se,Te) were determined by plotting the Nelson Riley function versus the lattice parameter calculated for each peak of the $\theta$-$2\theta$ scans[21]. The lattice parameter *a* was derived from reciprocal space maps (RSM) measured with a Panalytical X'pert Pro system with Cu-$K_\alpha$ radiation. Transmission electron microscopy (TEM) investigations of the films were performed in a FEI Tecnai-T20 TEM operated at 200 kV acceleration voltage. TEM lamellae were prepared by focused ion beam technique (FIB) in a FEI Helios 600i using an acceleration voltage of 3 kV in the last FIB step. The film thickness was confirmed to be 90 ± 5 nm. The composition of the samples was determined by energy-dispersive X-ray spectroscopy (EDX) with an Edax EDAMIII spectrometer. EDX linescans across the cross-section of the films confirm the stoichiometry to be homogeneous over the film thickness. It was found that the composition of the films is $FeSe_{0.7}Te_{0.3}$ for all studied substrates, irrespectively, if a sintered or single crystalline $FeSe_{0.5}Te_{0.5}$ target was used. The stoichiometry of the films might originate from a loss of Te during the PLD process due to its high vapor pressure and simultaneous Fe enrichment of the target surface due to the used low energy density of the laser beam. Electrical transport properties were measured in a Physical Property Measurement System [(PPMS) Quantum Design] by a standard four-probe method.

Figure 1 summarizes the structural characterization of the FeSe$_{0.7}$Te$_{0.3}$ films deposited on the different templates by XRD. In Fig. 1(a), only sharp 00$l$ peaks of the FeSe$_{0.7}$Te$_{0.3}$ films, the substrates and the CaF$_2$ buffer layers are present, indicating high phase purity with $c$-axis alignment of the films. The $\varphi$-scans of the 101 reflection for all FeSe$_{0.7}$Te$_{0.3}$ films, Fig. 1(b), show a fourfold symmetry, indicating that the films are epitaxially grown. Here, the respective epitaxial relation for the films on CaF$_2$, CaF$_2$-buffered LSAT and LSAT are identified as (001)[100]FeSe$_{0.7}$Te$_{0.3}$||(001)[110]CaF$_2$, (001)[100]FeSe$_{0.7}$Te$_{0.3}$||(001)[110]CaF$_2$||(001)[100]LSAT and (001)[100]FeSe$_{0.7}$Te$_{0.3}$||(001)[100]LSAT. However, the FeSe$_{0.7}$Te$_{0.3}$ on bare LSAT shows a significant larger full width at half maximum FWHM value, $\Delta\varphi$, of 8.12°, indicating that the crystalline quality of the film deposited on bare LSAT is inferior to the other films. On the other hand, the FeSe$_{0.7}$Te$_{0.3}$ film deposited on CaF$_2$-buffered LSAT has a $\Delta\varphi$ of around 0.7°, similar to the one on CaF$_2$ single crystals, regardless of CaF$_2$-buffer layer thickness.

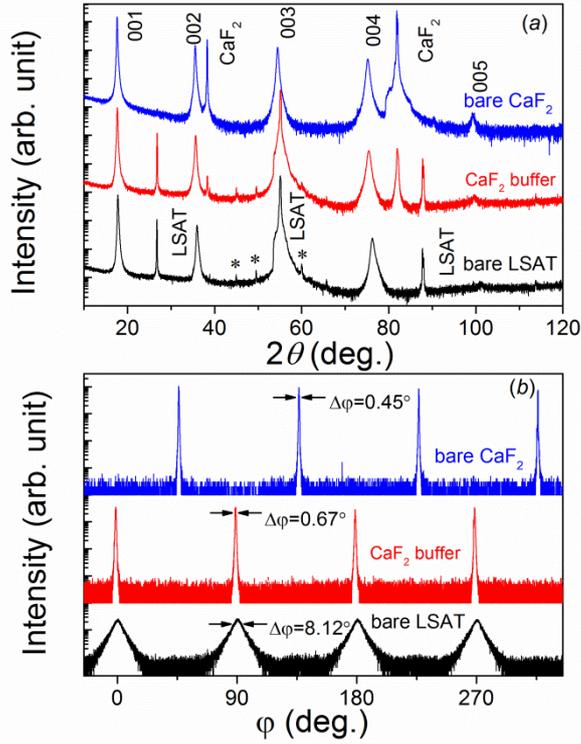

Fig. 1. XRD patterns for FeSe$_{0.7}$Te$_{0.3}$ films deposited on bare LSAT, CaF$_2$-buffered LSAT (80 nm of CaF$_2$ buffer) and bare CaF$_2$: (a) $\theta/2\theta$ scan and (b) $\varphi$-scans using the 101 reflection of FeSe$_{0.7}$Te$_{0.3}$. The asterisks mark reflections of the main substrate peak arising from secondary radiation of the x-ray tube.

The cross-sectional scanning TEM images of the FeSe$_{0.7}$Te$_{0.3}$ thin films on different substrates are presented in Fig. 2. The large area view of the film on bare LSAT is shown in Fig. 2(a). The areas with different contrast indicate different crystal orientations in the $ab$-plane, which is consistent with the XRD $\varphi$-scans. As shown in Fig. 2(b), the interface between LSAT and the FeSe$_{0.7}$Te$_{0.3}$ layer is clean. The FeSe$_{0.7}$Te$_{0.3}$ layers seem to contain no correlated defects except the in-plane grain boundaries mentioned above. A cross-sectional TEM image of the film on LSAT with $CaF_2$ buffer layer (80 nm) is shown in Fig. 2(c). A unique structure is observed at the interface between $FeSe_{0.7}Te_{0.3}$ and the $CaF_2$ buffer layer. Many triangular shaped features with several tens of nanometers in length are observed at the interface, which reflects the preferred habit plane for $CaF_2$ crystal (i.e., [111]). A similar observation is reported in Ref. 9. In spite of the rough interface between the $CaF_2$ buffer layer and $FeSe_{0.7}Te_{0.3}$, the thin films exhibit a flat surface. An area near the interface between the $FeSe_{0.7}Te_{0.3}$ film and the $CaF_2$ buffer layer is shown in Fig. 2(d). Surprisingly, the ordered atomic $FeSe_{0.7}Te_{0.3}$ planes immediately appear from the bottom of the valley of the $CaF_2$ buffer layer. Additionally, the $FeSe_{0.7}Te_{0.3}$ layer neither contains extended defects nor high angle grain boundaries. In contrast, no peculiar triangular features are found at the interface between $FeSe_{0.7}Te_{0.3}$ and $CaF_2$ substrate, as shown in Fig. 2(e). Nevertheless, a bright area of about 5 nm thick is observed at the interface between the $CaF_2$ substrate and the $FeSe_{0.7}Te_{0.3}$ film, similar to the observations in previous reports[22, 23]. Presumably this is a reaction layer between film and substrate. The sharp boundary between $FeSe_{0.7}Te_{0.3}$ and reaction layer indicates that the reaction layer is likely formed on the $CaF_2$ substrate side of the interface. The possible origin will be discussed below.

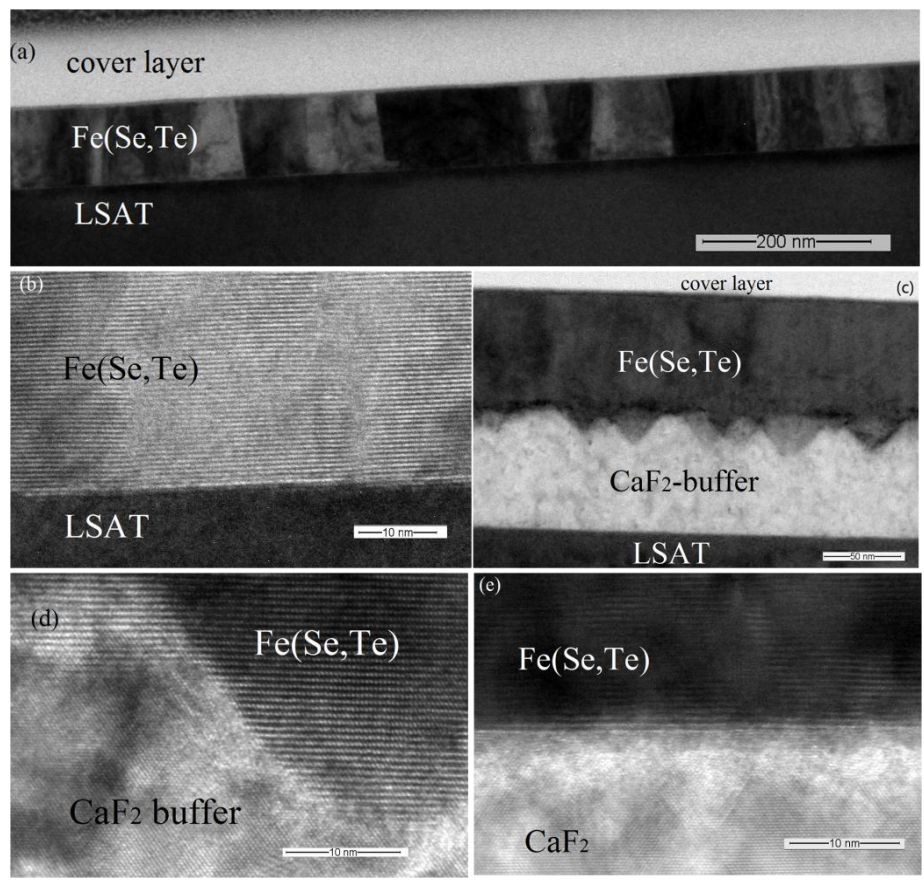

Fig. 2. Cross-sectional TEM images of the $FeSe_{0.7}Te_{0.3}$ thin film on: (a) bare LSAT and (b) interface between LSAT substrate and $FeSe_{0.7}Te_{0.3}$ film, (c) 80 nm thick $CaF_2$ buffered LSAT, (d) interface between $CaF_2$ buffer and $FeSe_{0.7}Te_{0.3}$ superconducting film, (e) interface between $CaF_2$ single crystal substrate and $FeSe_{0.7}Te_{0.3}$ superconducting film.

The analysis of the RSM measurements shows a clear dependence of the in-plane lattice parameter on the substrate used. The films on bare LSAT substrates have the largest $a$-axis lattice parameter, with a value of around 3.79 Å. This agrees well with the bulk lattice parameter of this material indicating a relaxed growth due to the high number of grain boundaries in this sample observed by the TEM[24]. By utilizing a CaF$_2$ buffer layer, the $a$-axis value decreases to 3.775 Å, irrespective of the fact that the in-plane lattice parameter of the CaF$_2$ buffer is $a/\sqrt{2} = 3.87$ Å almost identical to that of the LSAT substrate. Additionally, no influence of the CaF$_2$ buffer thickness on the $a$-axis length of the FeSe$_{0.7}$Te$_{0.3}$ films is observed. Finally, the FeSe$_{0.7}$Te$_{0.3}$ films on bare CaF$_2$ have the smallest $a$-axis length, ~3.75 Å. In general, such small $a$-axis lengths of the FeSe$_{0.7}$Te$_{0.3}$ films on bare CaF$_2$ substrates cannot be explained by the coherent epitaxial growth between the substrate and the film as the lattice parameter of CaF$_2$ is larger than the $a$-axis of FeSe$_{0.7}$Te$_{0.3}$. However, the coefficients of linear thermal expansion of the substrates are $18.9 \times 10^{-6}$ K$^{-1}$ for CaF$_2$[25], and $8.2 \times 10^{-6}$ K$^{-1}$ for LSAT[26], which are both higher than that of Fe(Se,Te)[27], having the value of $4.2 \times 10^{-6}$ K$^{-1}$. Consequently, the larger shrinkage of the substrate in comparison to the film leads to a compressive strain[12, 28, 29], during the cooling process after deposition resulting in a shorter $a$-axis length of the film. The expected change is larger for the film on CaF$_2$ single crystal in comparison to the LSAT-based films. We assume that this is the major contribution. Additionally, Ichinose *et al.* attributed the shrinkage of the $a$-axis lattice constant of Fe(Se,Te) film to an inter-diffusion of Se$^{2-}$ and F$^-$ at the interface[9]. This might give an additional contribution to the lattice parameter change and would explain the difference in the $a$-axis for the FeSe$_{0.7}$Te$_{0.3}$ films on LSAT and CaF$_2$-buffered LSAT, respectively. However, no clear evidence for Se diffusion was found in EDX experiments done in our TEM study of the interface between the FeSe$_{0.7}$Te$_{0.3}$ films and the CaF$_2$.

The temperature dependencies of the normalized resistance for the FeSe$_{0.7}$Te$_{0.3}$ films grown on the different substrates are shown in Fig. 3. All films show superconductivity, however, the normal state behavior is different. The film on bare LSAT exhibits a crossover from metallic to semiconducting-like behavior with decreasing temperature prior to the superconducting transition. The resistive upturn just before the transition is suppressed for the films on CaF$_2$ buffer layer (80 nm). The inset of Fig. 3 displays an enlarged view in the vicinity of the superconducting transition. $T_c$ is defined as 90% of the resistance in the normal state, which is just before the superconducting transition. The lowest $T_c$ of 4.9 K is measured for the film on bare LSAT with the poorest degree of texture. One of the most interesting features in Fig. 3 is the significant enhancement of $T_c$ by employing a CaF$_2$ buffer layer. $T_c$ is increased to 12.4 K for the film with CaF$_2$ buffer on LSAT. However, it is lower in comparison to the film on bare CaF$_2$. In this case, FeSe$_{0.7}$Te$_{0.3}$ has the highest $T_c$.

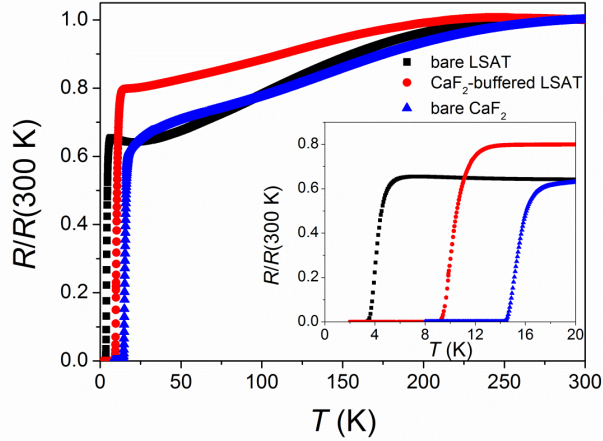

Fig. 3. Normalized resistive traces for FeSe$_{0.7}$Te$_{0.3}$ films on a variety of templates in zero magnetic field. The inset shows a close view on the superconducting transition.

In order to emphasize the correlation between the structural parameters and transport properties, we tentatively plotted $T_c$ as a function of lattice constants for a series of the FeSe$_{0.7}$Te$_{0.3}$ films on different substrates (Fig. 4). As shown in Fig. 4(a), the film on bare LSAT has the lowest $T_c$ with the largest $a$ value. Moreover, the film on bare CaF$_2$ has the highest $T_c$ with the shortest $a$-axis value. Despite the variation of the substrates, it is clear that the superconducting transition temperature $T_c$ is observed to decrease linearly with increasing $a$-axis lattice parameter for the FeSe$_{0.7}$Te$_{0.3}$ films, which is consistent with the previous report[30]. Therefore, the shrinkage of the $a$-axis lattice parameter, related to the compressive strain induced by the CaF$_2$ substrate due to the large thermal misfit and an additional inter-diffusion layer, might be crucial for such a high $T_c$ value. In contrast, the behavior of $T_c$ is widely scattered against the $c$-axis length, as shown in Fig. 4(b). Therefore, these results strongly suggest that the $a$-axis length is the dominant factor, which affects the superconducting properties of FeSe$_{0.7}$Te$_{0.3}$ films.

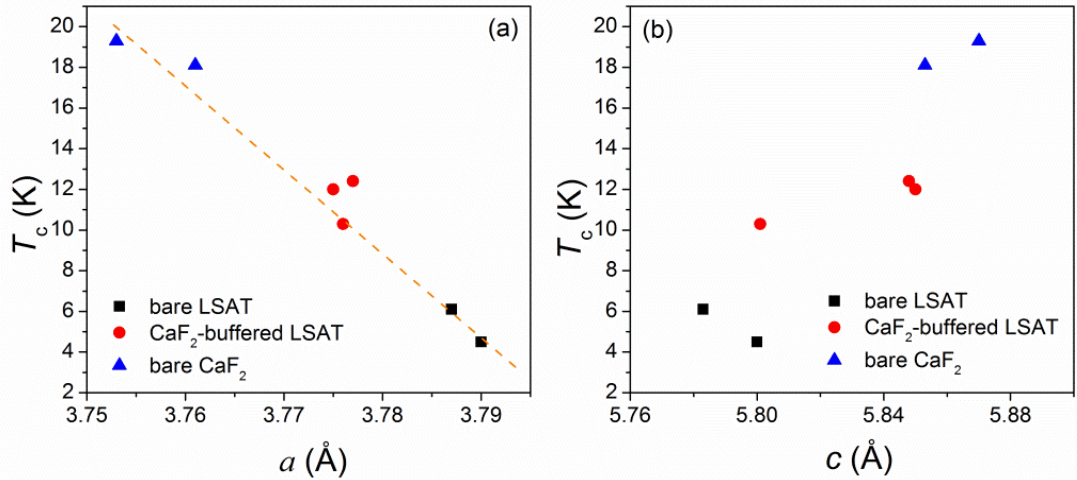

Fig. 4. The superconducting transition temperature, $T_c$ as a function of (a) $a$-axis and (b) $c$-axis lattice parameters. The line is a guide for the eye.

In summary, FeSe$_{0.7}$Te$_{0.3}$ thin films were deposited on a variety of templates by PLD. The film on bare LSAT has the lowest $T_c$ with the largest $a$ value. However, a CaF$_2$ buffer layer significantly improves the crystalline quality and superconducting properties of the FeSe$_{0.7}$Te$_{0.3}$ films. Furthermore, for the film on bare CaF$_2$, the $a$-axis lattice parameter shrinks due to the epitaxial compressive strain from CaF$_2$ originating from the thermal expansion mismatch. Thus, the in-plane lattice mismatch between Fe(Se,Te) and the substrates (CaF$_2$ and LSAT) is not the key factor for the $a$-axis lattice parameter of the film. Additionally, $T_c$ of FeSe$_{0.7}$Te$_{0.3}$ films is dominantly affected by the $a$-axis length.

The authors thank S. Richter (IFW Dresden) for fruitful discussions, J. Scheiter for help with TEM lamellae preparation, and M. Kühnel and U. Besold for technical support. The work was partly supported by the National Science Foundation of China (Grant No. NSFC-U1432135,11674054 and 11611140101) and Open Partnership Joint Projects of JSPS Bilateral Joint Research Projects (Grant No. 2716G8251b), the JSPS Grant-in-Aid for Scientific Research (B) Grant Number 16H04646 and the DFG funded GRK1621. V. G. is grateful to the DFG (GR 4667/1-1) for financial support. The publication of this article was fund by the Open Access Fund of the Leibniz Association.

10